\documentclass[nofigure]{pasj00}
\usepackage{graphicx}
\begin{document}
\SetRunningHead{L. V. T\'oth et al.}{L. V. T\'oth et al.}
%\Received{}%{yyyy/mm/dd}
%\Accepted{}%{yyyy/mm/dd}
%\Published{}%{yyyy/mm/dd}

\title{The AKARI FIS YSO Catalogue}

%%% begin:list of authors
% Do NOT capitalize all letters in "textsc".
\author{L. Viktor \textsc{T\'OTH} \altaffilmark{1,2},
G\'abor \textsc{MARTON} \altaffilmark{3,1},
Sarolta \textsc{ZAHORECZ}  \altaffilmark{1,2},
Lajos G. \textsc{BAL\'AZS} \altaffilmark{3,1},
Munetaka \textsc{UENO} \altaffilmark{4},
Motohide \textsc{TAMURA} \altaffilmark{5},
Akiko \textsc{KAWAMURA} \altaffilmark{6},
Zolt\'an T. \textsc{KISS},%, \altaffilmark{2}
Yoshimi \textsc{KITAMURA} \altaffilmark{7}}
%%% end:list of authors

%%% Please use the following style in case that sorting by
%%% affiliation is impossible.
%
% \author{%
%   D-Firstname \textsc{D-Familyname}\altaffilmark{1}
%   E-Firstname \textsc{E-Familyname}\altaffilmark{1,2}
%   and
%   F-Firstname \textsc{F-Familyname}\altaffilmark{2}}
\altaffiltext{1}{Department of Astronomy of the Lor\'and E\"otv\"os University,
             P\'azm\'any P\'eter s\'et\'any 1, 1117 Budapest, Hungary}
\email{l.v.toth@astro.elte.hu, marton.gabor@csfk.mta.hu, s.zahorecz@astro.elte.hu}
\altaffiltext{2}{Max-Planck-Institut f\"ur Astronomie, D-69117 Heidelberg, K\"onigstuhl 17}
\altaffiltext{3}{Konkoly Observatory of the Hungarian Academy of Sciences, H-1121 Budapest, Konkoly Thege Mikl\'os \'ut 15-17.}
\altaffiltext{4}{Japan Aerospace Exploration Agency (JAXA), 3-1-1 Yoshinodai, Sagamihara, Kanagawa 229-8510, Japan}
\altaffiltext{5}{National Astronomical Observatory of Japan, 2-21-1 Osawa, Mitaka, Tokyo 181-8588; The Graduate University for Advanced
Studies (SOKENDAI), 2-21-1 Osawa, Mitaka, Tokyo 181-8588}
\altaffiltext{6}{Department of Astrophysics, Nagoya University, Chikusa-ku, Nagoya 464-8602}
\altaffiltext{7}{Institute of Space and Astronautical Science, Japan Aerospace Exploration Agency, 3-1-1 Yoshinodai, Sagamihara, Kanagawa
229-8510}

%\email{eeeee@xxx.xxx.xx.xx}
%\altaffiltext{2}{Address of Institute}

%% `\KeyWords{}' always has to be placed before `\maketitle'.
 %Do NOT move this preamble from here! 

\maketitle

\begin{abstract}
We demonstrate the use of the AKARI survey photometric data in the study of galactic star formation. Our aim was to select young stellar objects (YSOs) in the AKARI FIS catalogue. We used AKARI Far-Infrared Surveyor and Wide-field Infrared Survey Explorer data to derive mid- and far-infrared colours of YSOs. Classification schemes based on Quadratic Discriminant Analysis have been given for YSOs. The training catalogue for QDA was the whole sky selection of previously known YSOs (i.e. listed in SIMBAD). A new catalogue of AKARI FIS YSO candidates including 44001 sources has been prepared. Reliability of the classification is over 90\% as tested in comparison to known YSOs. As much as 76\% of our YSO candidates are from previously uncatalogued type. The vast majority of these sources are Class I and II types according to the Lada classification. The distribution of AKARI FIS YSOs' is well correlated with that of the galactic ISM. Local over densities were found on infrared loops and towards the cold clumps detected by Planck. 
\end{abstract}

\section{Introduction}

%\noindent IMPORTANT NOTICE\\
%1. ``\verb|\draft|'' creates single column and double spaces format.\\
%2. If you comment out ``\verb|\draft|'', the output will be double column and single space.\\
%3. For cross-references, the use of ``\verb|\label|, \verb|\ref|, \verb|\cite|'' and the thebibliography environment is strongly recommended. \\
%4. Do NOT use ``\verb|\def|, \verb|\renewcommand|''.\\
%5. Do NOT redefine commands provided by PASJ00.cls.\\
The AKARI FIS BSC (Yamamura et al. 2010) is currently the only public unbiased all sky point source catalogue beyond $100\,\mu$m. In this paper we examine the possible benefits of it in the study of young stellar objects (YSOs). 
Unbiased surveys covering large fields and large number of objects provide us with the possibility to study star formation and early stellar evolution. T Tauri stars were discovered well before the 1980s as bright optical variables with G, K, or M spectral types and strong H\,I and Ca\,II emission lines (Joy 1946, 1949). 
Detailed studies of these objects however require infrared measurements as well.
Large area surveys of the infrared sky were first carried out forty years ago, using ground based (Neugebauer et al. 1969) and baloon born (Hoffman et al. 1967) instruments. 
These first surveys revealed candidates for very early phases of the stellar evolution (see e.g. the BN object, Becklin and Neugebauer, 1967). 
The observed infrared (IR) emission originates predominantly from the contracting gaseous envelopes and the circumstellar disks in the dusty environment of the forming stars. 
While infrared measurement remains the number one tool in exploring the circumstellar medium, modelling efforts of the young stellar systems are tracing their radiation on all wavelengths from the X-ray (see e.g. Walter et al. 1987) to the microwave (see e.g. Bieging \& Cohen, 1985) regime.
Nearby low mass YSOs are quite numerous due to the shape of the stellar IMF and the slower evolution of the less massive YSOs. 
Definition of evolutionary stages, and refinement of the YSO models have been made possible by surveys like IRAS (Beichman et al. 1985). 
The IR spectral energy distributions (SEDs) have been successfully used to model the composition and geometry of the circumstellar disks and of the envelopes. 
Large area infrared surveys implied simple identification of stars in formation and pre-main sequence (PMS) evolution via their colours. 
Object types like ''T Tau'', ''cold cloud core'', ''stellar'' or ''galaxy like'' were identified in colour-colour diagrams drawn from [25/12], [60/25] and [100/60] IRAS colours (see Emerson 1987, Myers et al. 1987), also classical T Tau type point sources in the 2MASS (Cutri et al. 2003) J-H, H-K colour-colour diagram.
The YSOs were classified according to their evolutionary phases (Figure \ref{atlagsed}; Lada 1987; Adams et al. 1987; Adams $\&$ Shu 1988; Pezzuto et al. 2002, ISOLWS).
Embedded sources, sometimes known as protostars, are optically invisible YSOs with SEDs peaking at mid-infrared (MIR) to far-infrared (FIR) wavelengths. Classical T Tauri stars (CTTs) have strong emission lines and substantial IR or UV excesses. 
Weak emission T Tauri stars have weak or no emission lines and negligible IR excesses. 
These objects form a rough age sequence with protostars as the youngest stellar objects and weak emission T Tauri stars as the oldest. 
ISO (Franceschini et al. 1995) and Spitzer Space Telescope (SST, Werner et al. 2004) surveys allowed a critical review of theories of the composition and structure of dust grains.
As a new approach, Stage I and Stage II classes were defined and separated by Robitaille (2006) in SST MIR two colour diagrams.
Answers to the open questions of star formation and PMS evolution require a large number of sources which enable a better coverage of the parameter space. 
Recently Evans at al. (2009) questioned the time scales of YSO evolutionary phases based on a large area SST survey covering the nearby star forming regions Perseus and Ophiucus. Assuming a continuous flow through the YSO classes, they derive median lifetimes for prestellar, Class 0, Class I and Class II  phases of 0.46, 0.16, 0.54 and 0.40 Myr, respectively.

The PMS evolution is largely an evolution of the circumstellar matter. 
Proto-planetary disks are known as the transition between the protostellar phase and planet formation, nevertheless not all of them may form planets. 
It is crucial to measure the emission of the cold part of the disk i.e. at around 100\,$\mu $m and beyond 100\,$\mu $m wavelengths in a survey. The IRAS Point Source Catalogue (Beichman et al. 1988) contained the first uniform all sky FIR photometry of YSOs. The AKARI Far Infrared Surveyor (FIS) all sky survey may be considered as a "super IRAS" with an increased sensitivity and resolution and with observations also beyond 100$\mu $m.
Finding an empirical relationship connecting physical properties of interstellar gas and star formation activity is drawing much attention since the work of Schmidt (1959). Massive nearby clouds with relatively low star formation activity suggest that yield and rate of star formation can vary considerably in molecular clouds, independent of their mass and size - however, there exists a linear relationship between star formation rate and extinction (Lada et al. 2010).
We present a classification based on Quadratic Discriminant Analysis (QDA) separating YSO candidates according to AKARI FIS photometry. In addition, we compare galactic ISM structures by investigating their galactic distribution. 
The accretion process may depend on the physical conditions of the environment, that may be tested checking the ISM around the YSOs (see e.g. recently by Kirk \& Myers 2010).

\section{Input data and data analysis }

\subsection{AKARI data}
\begin{figure}[h]
	\centering
	\includegraphics[width=9cm]{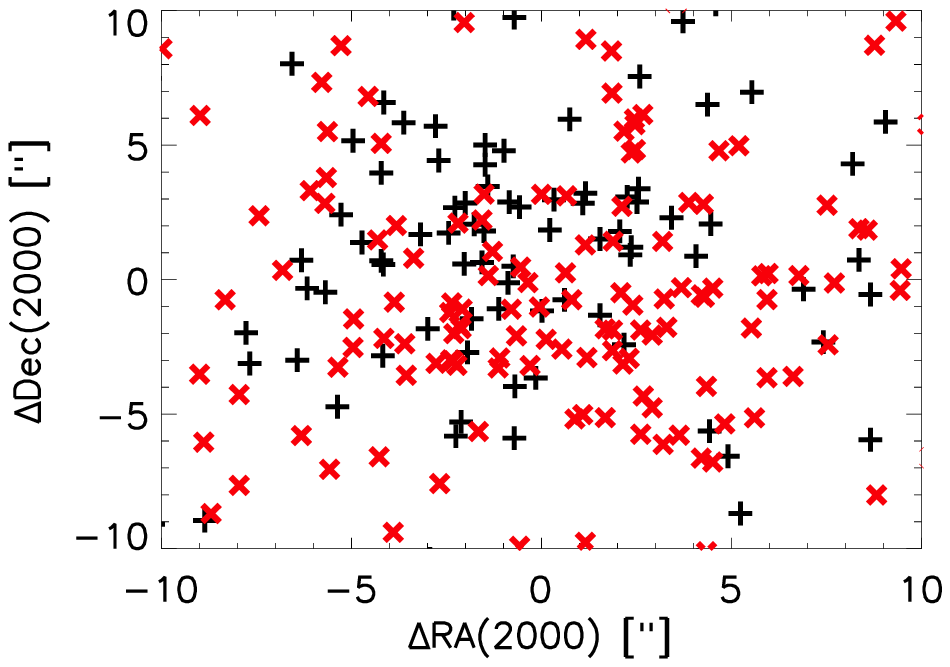}
	\caption{Positional difference for selected AKARI FIS BSC sources (see text). Black pluses show the positional difference between AKARI FIS BSC point sources and 2MASS point sources. Red crosses show the positional difference between AKARI FIS BSC point sources and WISE point sources. The mean positional difference is 7.18'' and the standard deviation is 4.15$^{\prime \prime}$. \label{pointing}}
\end{figure}

The AKARI Japanese satellite for infrared astronomical purposes operated with a telescope of 68.5 cm in diameter (Murakami et al. 2007). The telescope and focal plane instruments were cooled to a temperature lower than 6 K in a liquid-helium cryostat (Nakagawa et al. 2007) to avoid the thermal emission of the instruments. The FIS is one of the two focal plane instruments onboard (Kawada et al. 2007). Four photometric bands were used to scan the infrared sky between 50 and 180\,$\mu$m. Detectors and bands are the following: N60 (65\,$\mu$m), WIDE-S (90\,$\mu$m), WIDE-L (140\,$\mu$m), N160 (160\,$\mu$m). 
The AKARI FIS Bright Source Catalogue (BSC, Yamamura et al. 2010) lists 427071 point sources detected at least in one of the four FIS bands. 
Calibration and characterisation of FIS were performed by observing well-established photometric calibration standards, like solar system-objects and stars. These sources have been studied extensively (M\"uller $\&$ Lagerros 1998, 2002, Cohen et al. 1999, 2003a, 2003b) and were used as calibration sources in earlier infrared missions (e.g., ISO; Blommaert et al. 2003, Schulz et al. 2002, COBE; Hauser et al. 1998).\\
The quality of the flux densities is given with quality flags: "1" for upper limits, "2" for moderate quality and "3" for good quality (see Yamamura et al. 2010). There are as many as 11543 point sources with four good or moderate flux qualities, while 21929 has at least 3, and 73962 has at least 2 good or moderate flux qualities. The typical flux uncertainty is 10 $\%$.\\
The positions of the AKARI FIS BSC point sources are given in J2000 equatorial coordinates with an assumed positional accuracy of 8 arcseconds (Yamamura et al. 2010). The pointing accuracy was tested comparing 2MASS, WISE and AKARI coordinates. For this test we used 160 point sources in the Tau-Aur-Per (hereafter TAP) region having at least two good flux qualities and associated YSO or infrared point source counterparts in the Simbad database. Coordinate differences of associated 2MASS, WISE and AKARI point sources are shown in Fig.~\ref{pointing}. The average pointing error is around 7 arcseconds.

\subsection{WISE data}

The Wide-field Infrared Survey Explorer (WISE, Wright et al. 2010) operates with a 40 cm diameter telescope, including six mirrors before the scan mirror and six mirrors in the camera after the scan mirror. The cryostat uses solid hydrogen to cool the telescope to less than 12.5\,K and the Si:As detector arrays to less than 7.5 K. Until the end of the hydrogen WISE mapped the whole sky in four bands centered at 3.4, 4.6, 12 and 22 $\mu m$. In unconfused regions WISE is achieving 5$\sigma$ point source sensitivities better than 0.08, 0.11, 1 and 6 mJy with angular resolution 6.1, 6.4, 6.5 and 12.0 arcseconds, respectively. The astrometric precision is better than 0.15 arcseconds for the sources with a high signal-to-noise ratio. 
The WISE All-Sky Source Catalog (Cutri et al. 2012) lists 563921584 sources. We used fluxes of associated WISE point sources with photometric uncertainties lower than 0.2 magnitudes in all 4 bands.

\subsection{Other data}\label{otherdata}

Besides the AKARI, 2MASS \& WISE we used the following point source catalogs: optical data from USNO-B1 catalogue (Monet et al. 2003); infrared data from the Spitzer Space Telescope Archive, submm data from SCUBA, SHARCII (Andrews $\&$ Williams, 2005) and CSO (Beckwith $\&$ Sargent 1991).

The distribution of ISM was analysed using the CO line intensity map from Dame et al. (2001) large-scale CO survey obtained with 1.2 m telescopes (HPBW=8'). Also, the dust maps of Schlegel et al. (1998) were used in a detailed analysis of galactic structures.

\subsection{Spectral response functions}
\begin{figure}[h]
  	\centering
	\includegraphics[width=9cm]{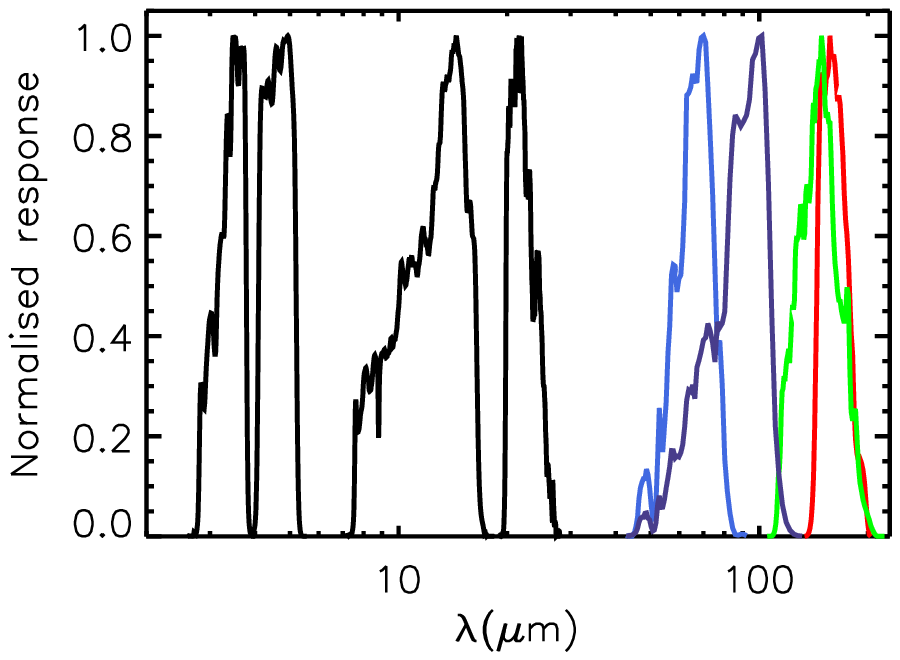}
    \caption{Normalised spectral response functions of the WISE (black curves) and AKARI FIS (coloured curves) filters. We note that the AKARI FIS filters are overlapping.} \label{responses}
\end{figure}

The AKARI and WISE normalised response functions of the detector-filter systems are plotted in Fig. \ref{responses} from Kawada et al. (2007) and Wright et al. (2010). In Table \ref{szurok} the basic parameters of the response functions are listed: $\lambda_{nom}$ is the central wavelength in $\mu m$, $\Delta\lambda_{eff}$ is the bandwidth in $\mu m$ and FWHM in $\mu m$. The WISE and AKARI FIS filters cover well the MIR and FIR wavelength range.

\begin{table}[h]
\caption{Properties of the infrared filters}
\label{szurok}
\centering
\begin{tabular}{c r r r}
\hline
Filter name & $\lambda_{nom}$ & $\Delta\lambda_{eff}$& FWHM \\
 &  $[\mu m]$ & \ $[\mu m]$ &$[\mu m]$\\
\hline\hline
WISE Band 1 & 3.35 & 0.66 & 0.65\\
WISE Band 2 & 4.60& 1.04 & 1.11\\
WISE Band 3 & 11.56& 5.51 & 6.28\\
WISE Band 4 & 22.09& 4.10 & 4.74\\
AKARI N60 & 65 & 21.70  & 18.26 \\
AKARI WIDE-S& 90 & 37.90  & 24.46\\
AKARI WIDE-L & 140 & 52.40  & 37.47\\
AKARI N160 & 160 & 34.10 & 31.15 \\
\hline
\end{tabular}
\end{table}

\subsection{Comparison of modelled and observed AKARI FIS fluxes} \label{comparison}
\begin{figure}[h]
   \centering
	\includegraphics[width=9cm]{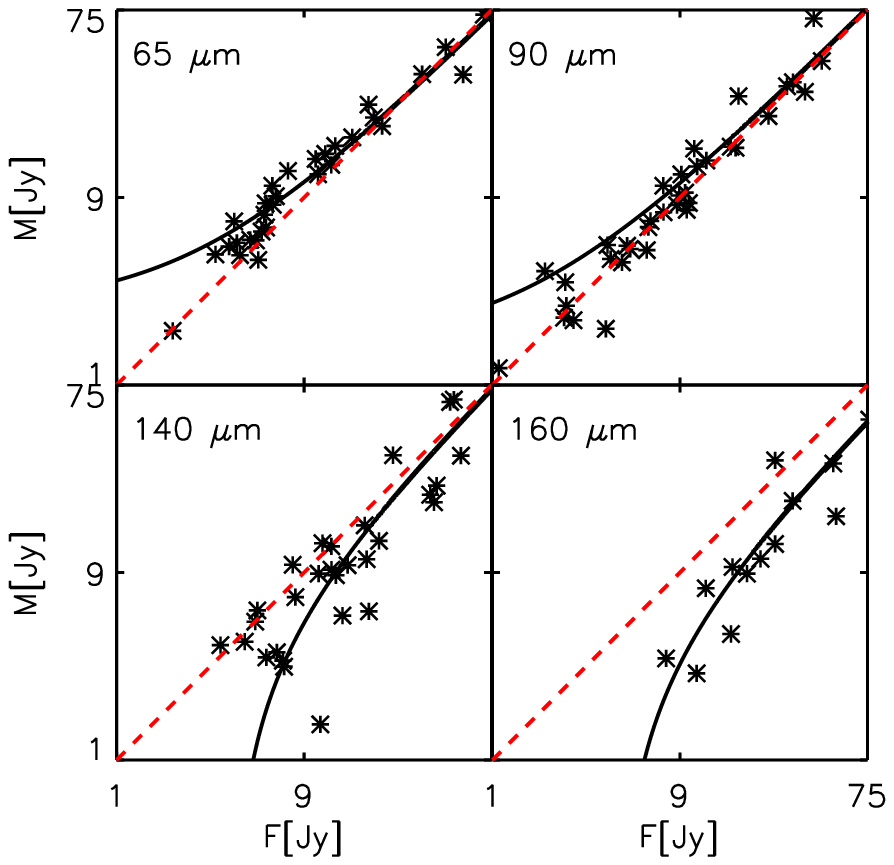}
	\vspace{0.5 cm}
\caption{Observed and calculated flux densities (see text) for the modelled YSOs. Red dotted lines mark the 1:1 ratio. Black solid lines mark the fitted lines. At 160 $\mu$m we found the following relation: $M[Jy]=0.69*F[Jy]-2.98$.
              \label{oc}}
\end{figure}

We made a detailed analysis of the 585 AKARI FIS BSC point sources in TAP region. SEDs were plotted using AKARI and other photometric data (see \ref{otherdata}). The stellar parameters of candidate YSOs were estimated using the online SED Fitting Tool of Robitaille et al. (2007)\footnote{http://caravan.astro.wisc.edu/protostars}. We have succeeded fitting model SEDs for 38 AKARI FIS BSC point sources in the TAP region. 
Calculated flux densities were derived from the best fits using the spectral response functions (see Table~\ref{szurok}). The correlations between calculated and observed fluxes are shown in Fig. \ref{oc}.
We conclude that the good quality AKARI FIS BSC flux densities at 65, 90 and 140 $\mu$m are well fitted into the SEDs of the known YSOs in TAP region. We found however a systematic deviation in the value of the 160 $\mu$m observed and modelled flux densities. Possible reasons could be calibrational uncertainties or inaccurate modelling of the far-infared part of the SED. In the AKARI bands the observed flux is dominated by the radiation of the circumstellar disk and envelope. Underestimation of the surrounding dust or warmer dust temperature can undervalue the FIR flux. We worked with precomputed YSO models, individual radiative transfer modelling of the sources can solve this problem.

The calculated flux densities are smaller than the FIS observed values. The difference is increasing with the flux density. The relationship between the observed and modelled flux densities at 160 $\mu m$ are:
\begin{equation}
M[Jy]=0.69*F[Jy]-2.98,
\end{equation}
where M is the modelled and F is the observed flux density.\\

\begin{figure}[h]
   \centering
	\includegraphics[width=9cm]{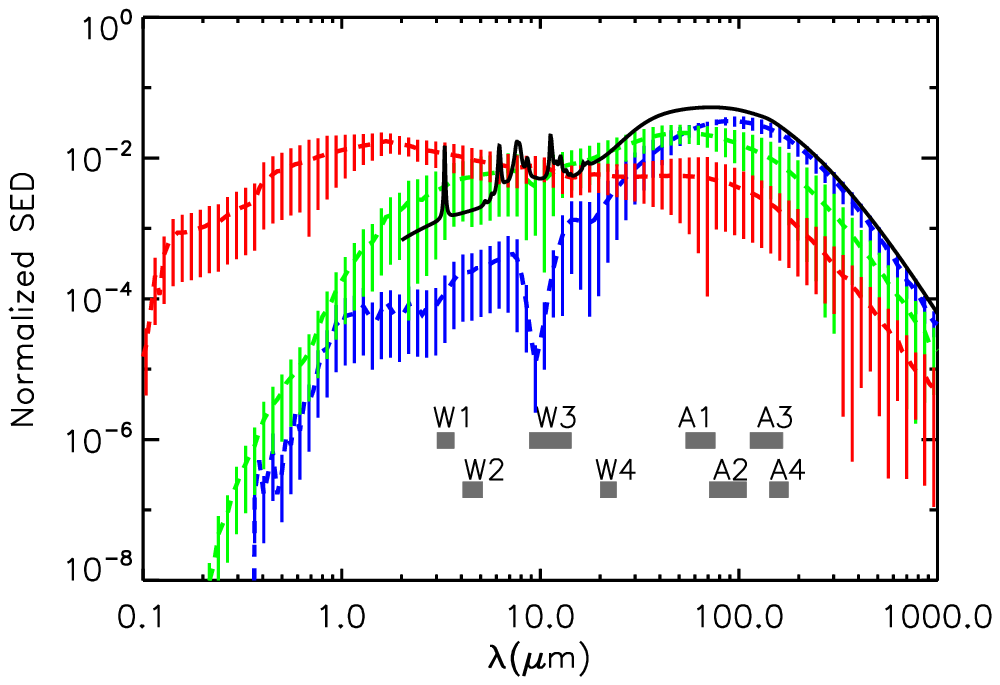}
   \caption{Normalised average SED in the TAP region of the Class 0 (blue), Class I (green) and Class II (red) objects. Black continuous line represents the typical SED of starburst galaxies. Grey rectangulares show the WISE and AKARI FIS bands. Great similarity is observable between the YSOs' and starburst galaxies' SEDs.}
              \label{atlagsed}
\end{figure}

\section{Classification of point sources}

YSOs are divided to Class 0, Class I and Class II based on their SEDs. Using the modelled YSOs, we produced normalised average SEDs of these three YSO groups in the TAP region (see Fig. \ref{atlagsed}). On Figure \ref{atlagsed} a starburst galaxiesÕ SED is also overplotted, a great similarity is observable between the Class I's and starburst galaxies' SED. Longward of 10$\mu$m the YSO's SED is dominated by the reprocessed flux of absorbed stellar radiation by the circumstellar dust. There is not a significant difference in the FIR flux of the Class 0 - Class II objects, only a small decrease is noticeable. Shortward of 10 $\mu$m the flux is dominated by stellar light and the radiation of warm dust in the inner disk region. The fraction of the stellar radiation is growing from Class 0 to Class II objects and they become observable at shorter wavelengths. The main difference between the different classes is the slope of the SEDs.

Our goal is to separate the different types of galactic objects and the extragalactic sources. We tried to estimate the average colours and average fluxes of selected types of objects. The separation was based on colour-colour and flux-colour diagrams. The WISE Catalog has been recently released. We include the WISE W1 magnitude and W1-W2 colour in our classification basis. In order to do this we looked for the associated sources in the WISE database using a searching radius of 5$^{\prime \prime}$. WISE objects nearest to the AKARI FIS sources with an error $<$ 0.2 mag in the W1 and W2 bands were then assigned to those AKARI objects having the highest quality flag in the two wide bands (W90 and W140). 

\subsection{Quadratic Discriminant Analysis}

For classification and pattern recognition in multi-dimensional data, one can use several statistical methods. Discriminant analysis (DA) is commonly used, because DA aims to make difference between groups in the multivariate parameter space, orders membership probabilities to the cases. One may use this scheme for classifying additional items not having assigned group membership. We used Quadratic Discriminant Analysis (QDA, McLachlan 1992) to classify the objects, because generally QDA offers increased flexibility over Linear Discriminant Analysis (Everitt et al. 2001). There are several approaches to solve this problem. These are usually among the major ingredients of the professional statistical software packages. We used R\footnote{http://www.r-project.org/} in our computations.

In order to separate the different types of sources firstly a training sample has to be prepared. This training sample contains the information which is the basis of the differentiation. QDA looks for quadratic boundaries between groups in the multi-dimensional parameter space. After determining the boundaries, the algorithm applies them to the requested data and orders membership probabilities for each source and each group. These probabilities (hereafter $\eta$) can then be used to determine the type of the source.

\begin{figure}[h]
   \centering
	\includegraphics[width=9cm]{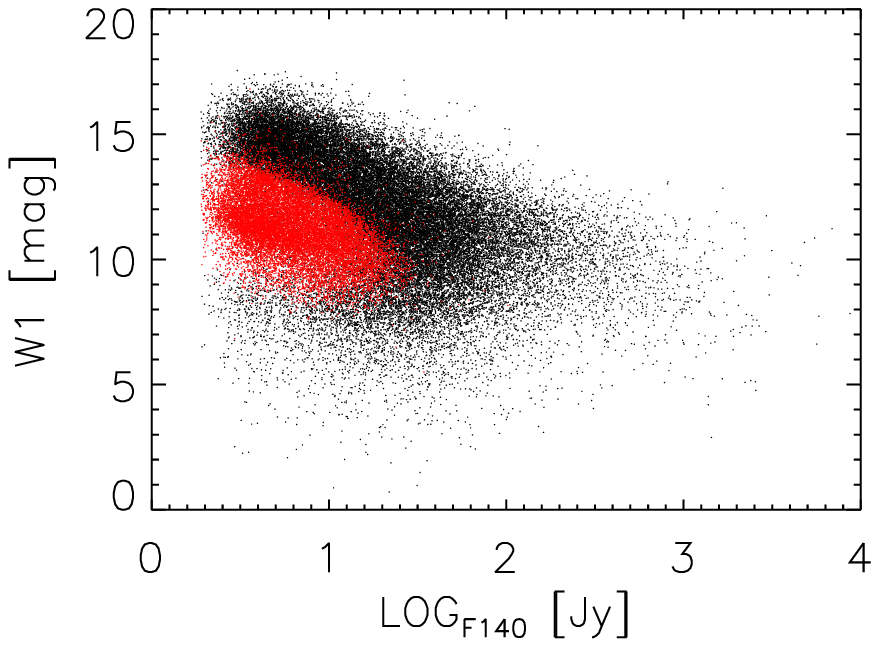}
   \caption{Slice of the multi-dimensional colour-colour and colour-flux space demonstrating that group boundaries are non-linear.}
              \label{szinszin}
\end{figure}

\subsection{Application of QDA to our data} \label{qda}

In the first step we made a simple position correlation between the FIS and SIMBAD catalogue using a 30$^{\prime \prime}$ search radius, which corresponds to the HPBW of the AKARI N160 filter PSF. In several cases not only one, but multiple SIMBAD sources were found within the above mentioned radius. For these cases all the associates were stored. However we note, that in SIMBAD each object has a main type defined and several other subtypes\footnote{http://simbad.u-strasbg.fr/simbad/sim-help?Page=sim-fsam\#Sotypes} . We dealt only with the main type of objects. We found association in 21736 cases for given AKARI FIS point sources and in 404406 cases there were no associated objects within 30$^{\prime \prime}$. After associating these sources to the WISE catalog, 64653 objects remained for further studies.

In the teaching phase we made the following assumptions: 1) the AKARI detections can be divided  into three major groups, namely galaxies, evolved (red) stars and YSOs; 2) galaxies are the objects having SIMBAD counterparts with any ÒGalaxyÓ object type and are located above $|b|=3$ degree; 3) red, evolved stars are the objects marked in SIMBAD as RG* (red giant branch star), AB* (asymptotic giant branch star), C* (carbon star), S* (S star), sg* (evolved supergiant star) or pA*(post-AGB star); 4) YSOs are the ones with object type Y*O (young stellar object), Y*? (young stellar object candidate), TT* (T Tau star), TT? (T Tau star candidate), Or* (variable star of Orion type), HH (Herbig-Haro object), pr* (pre-main sequence star), pr? (pre-main sequence star candidate), FU* (variable star of FU Ori type), HII (HII region), Em* (emission-line star), *iC (star in cluster) or WR* (Wolf-Rayet star). This way we were able to execute the QDA by using 4609 objects as galaxies, 2021 objects as YSOs and 13 objects as evolved stars.

In our case the classification was based on the logarithmic values of fluxes and colours, namely [F140] flux values, W1 magnitudes and [F65/F90], [F90/F140], [F140/F160] and W1-W2 colours (see Fig. \ref{szinszin}).

\section{Results and Discussion}\label{results}

From the aforementioned 64653 sources 44588 were found to be YSO with $\eta > 0.5$, 18494 were classified as galaxy-like object with $\eta>0.5$ and 1303 sources were found to be evolved star with $\eta>0.5$. 268 objects remained unclassified because they had all the $\eta<0.5$. 

\subsection{Reliability test of classification}

The reliability of our classification was tested by checking the SIMBAD object types previously assigned to the sources. We inspected the nearest associations within a distance of 30$^{\prime \prime}$ and also all the objects within that radius. Results are listed in Table ~\ref{assoc}. Columns are as follows: 1) Type as we classified with QDA; 2) Number of sources; 3) highlights if we are dealing with the nearest or all associated objects within 30$^{\prime \prime}$; 4) total number of first or all SIMBAD associations; 5) number of SIMBAD YSO type objects (defined as described in Sect. ~\ref{qda}); 6) number of SIMBAD IR type objects; 7) number of ISM type SIMBAD objects (this includes all types of objects belonging to the ISM main type); 8) SIMBAD "*" type objects (i.e. stars); 9) galaxies (defined as described in Sect. ~\ref{qda}); 10) evolved stars (as described also in Sect. ~\ref{qda}). Columns from 11-17 are the proportion of the total SIMBAD associates with given type.

The group of AKARI objects classified as YSOs contain the most SIMBAD YSOs, as we mentioned in Sect. ~\ref{comparison}. We found 1034 SIMABD YSOs, as closest associates for the classified point sources. 965 of them (93.3\%) were classified as YSO candidates with QDA. This value of accuracy is 93.9\% when considering all the SIMBAD associates within 30$^{\prime \prime}$. 

Our detailed analysis of all the 585 AKARI FIS BSC point sources in the TAP region (see \ref{comparison}) located 174 sources with YSO type SEDs. 68 of these got only one reliable FIS flux, therefore these were not included in our statistical selection. From the remaining 106 point sources our QDA selection classified 97 as YSOs (92\%). 7 of them were found to be a galaxy and 2 of them were classified as evolved star. We consider this as very high reliability of our statistical classification. We note that we located 184 YSOs with QDA in TAP. SED based classification was not possible for 78 (42\%) because of multiplicity or lack of quality flux data. We found only one galaxy among the QDA selected YSOs in the TAP region. 

The obvious contamination in the whole-sky QDA YSO sample caused by known galaxies is 587 (5.2\%) among the closest associates, while it is only 2.8\% when considering all the associates in 30$^{\prime \prime}$ vicinity of the sources. To prepare the final catalogue of YSO candidates, we removed those from the 44588 sources where the closest associate is a known galaxy in the SIMBAD database. This resulted in a final list of 44001 YSO candidate sources.\\

\begin{table*}[!ht]
\caption{Table of associated objects for QDA classified FIS sources as a result of position correlation with SIMBAD database. The number of closest associates and the number of all associated within 30$^{\prime \prime}$ vicinity are listed along with the percentage that they represent in the sample. For detailed explanation see text.}
\label{assoc}
\centering
\small\addtolength{\tabcolsep}{-5pt}
\begin{tabular}{cc|lcccccccc|cccccccc}
\hline
QDA type&Number&\multicolumn{8}{c}{SIMBAD associations}&\multicolumn{8}{c}{\% of total associations}\\
\hline
\hline
& & &Total& YSO & IR &mm+smm& ISM & * & G & EVO & YSO & IR &mm+smm& ISM & * & G & EVO\\
\hline
YSO & 44588 & nearest in $r<30^{\prime \prime}$ &11235 & 965 & 5254 & 1008 & 1027 & 724 & 587 & 55 & 8.6 & 46.8 & 9.0 & 9.1 & 6.4 & 5.2 & 0.5\\
  &   & all in $r<30^{\prime \prime}$ &28759 & 2295 & 7117 & 1905 & 4480 & 4175 & 810 & 93 & 8.0 & 24.7 & 6.6 & 15.6 & 14.5 & 2.8 & 0.3\\
 \hline
Galaxy-like & 18494 & nearest in $r<30^{\prime \prime}$ & 6602 & 63 & 1062 & 7 & 114 & 212 & 4564 & 5 & 1.0 & 16.1 & 0.1 & 1.7 & 3.2 & 69.1 & 0.1\\
  &   & all in $r<30^{\prime \prime}$ &14045 & 131 & 1502 & 20 & 1888 & 662 & 6098 & 13 & 0.9 & 10.7 & 0.1 & 13.4 & 4.7 & 43.4 & 0.1\\
 \hline
Evolved star & 1303 & nearest in $r<30^{\prime \prime}$ & 271 & 6 & 124 & 1 & 5 & 30 & 6 & 11 & 2.2 & 45.8 & 0.4 & 1.8 & 11.1 & 2.2 & 4.1\\
  &   & all in $r<30^{\prime \prime}$ &350 & 17 & 128 & 1 & 14 & 39 & 7 & 12 & 4.9 & 36.6 & 0.3 & 4.0 & 11.1 & 2.0 & 3.4\\
\hline
Not class.& 268 & nearest in $r<30^{\prime \prime}$ &23 & 0 & 14 & 0 & 0 & 6 & 1 & 0 & 0.0 & 60.9 & 0.0 & 0.0 & 26.1 & 4.3 & 0.0\\
with good qual.  &   & all in $r<30^{\prime \prime}$ &40 & 1 & 14 & 0 & 0 & 19 & 2 & 0 & 2.5 & 35.0 & 0.0 & 0.0 & 47.5 & 5.0 & 0.0\\
\hline
Other FIS& 357835 & nearest in $r<30^{\prime \prime}$ &54216 & 695 & 14820 & 708 & 2089 & 4575 & 21609 & 784& 1.3 & 27.3 & 1.3 & 3.9 & 8.4 & 39.9 & 1.4\\
detected  &   & all in $r<30^{\prime \prime}$ &79688 & 1508 & 19289 & 1201 & 4230 & 7992 & 26244 & 926 & 1.9 & 24.2 & 1.5 & 5.3 &10.0 & 32.9 & 1.2\\
\hline

\end{tabular}
\end{table*}

\subsection{The YSO catalogue}

Our catalogue built from the 44001 YSO candidates is available in electronic form at CDS. The catalogue contains the AKARI FIS IDs,  RA (2000), Dec (2000) celestial coordinates, WISE magnitudes at 3.6, 4.6, 12 and 22$\mu$m with the corresponding uncertainties, the AKARI FIS flux values and their photometric quality flags. For each source we list the probability of YSO candidate membership, and additional information based on SIMBAD: the main type of the nearest SIMBAD associate, its other object types and the SIMBAD ID of the source.

\subsection{The galactic distribution of YSO candidates}

\begin{figure}[!ht]
   \centering
	\includegraphics[width=9cm]{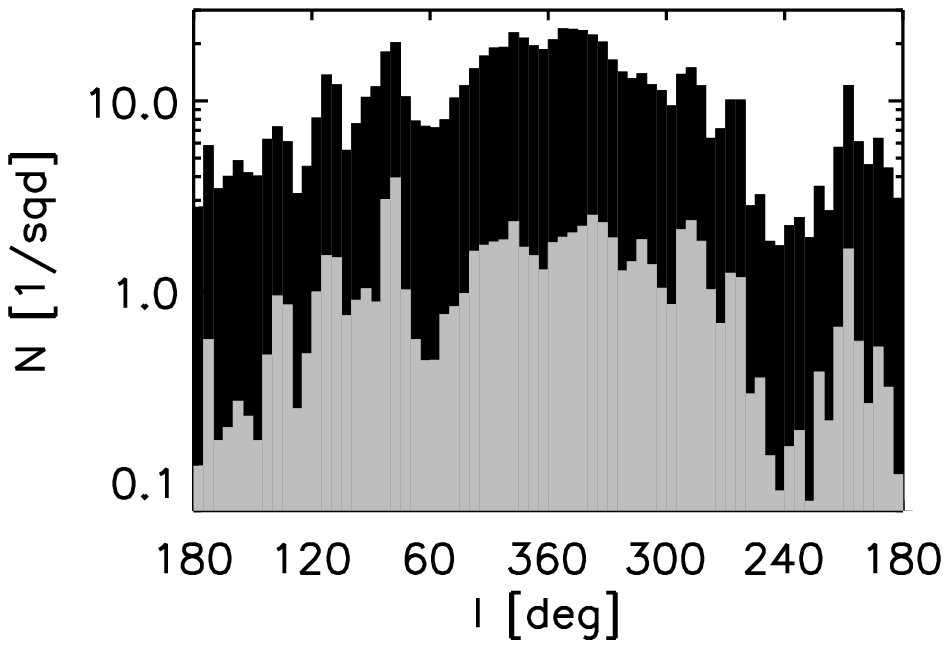}
   \caption{Surface density distribution of all the AKARI FIS BSC point sources (black) and of the 44001 YSO candidates (gray) along galactic longitude. Both distributions show maxima at $l\approx$0$^\circ$ and $l\approx80^\circ$, minimum at $l\approx240^\circ$.}
              \label{l_hist}
\end{figure}

\begin{figure}[!ht]
   \centering
	\includegraphics[width=9cm]{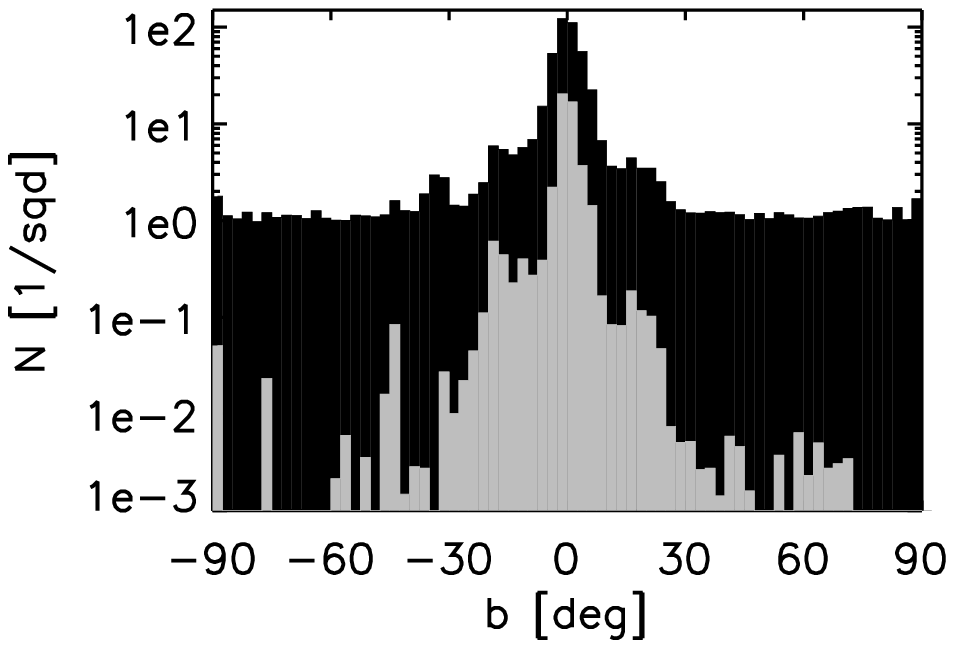}
   \caption{Surface density distribution of all the AKARI FIS BSC point sources (black) and of the 44001 YSO candidates (gray) along galactic latitude show a maximum at the galactic midplane. A local maximum is seen for QDA YSOs at b=-45$^\circ$. The southern galactic hemisphere shows a slight excess (53\%-46\%) in the number of YSO candidate objects.}
              \label{b_hist}
\end{figure}

We studied the surface density (hereafter $N$ [1/sqd]) distribution of YSO candidates found by the QDA, see Fig. \ref{l_hist} and Fig. \ref{b_hist}. The AKARI FIS point source distribution peaks at the galactic midplane, and there are surface density variations along the plane with maximum towards the Galactic Centre, a local maximum at $l=80^\circ$ and minimum at around $l=240^\circ$.
	 
The distribution of the 44001 YSO candidates shows similarities to the complete FIS point source catalogue with maximum at $l=80^\circ$ (Cyg-X SFR) and a minimum at $l=240^\circ$. Local maxima are in the direction of Orion  SFR ($l=200^\circ-210^\circ$), the Vela ($l=270^\circ$), the Chamaeleon ($l=290^\circ$), the Cepheus ($l=105^\circ$) and the Taurus-Auriga-Perseus region ($l=160^\circ-180^\circ$). A prominent maximum in the surface density distribution along the galactic latitude is present in the direction of the galactic midplane ($b=0^\circ$).\\

\subsubsection{The relative YSO surface density}

\begin{figure*}[!ht]
   \centering
	\includegraphics[trim=0cm 0cm 0cm 6cm width=\textwidth]{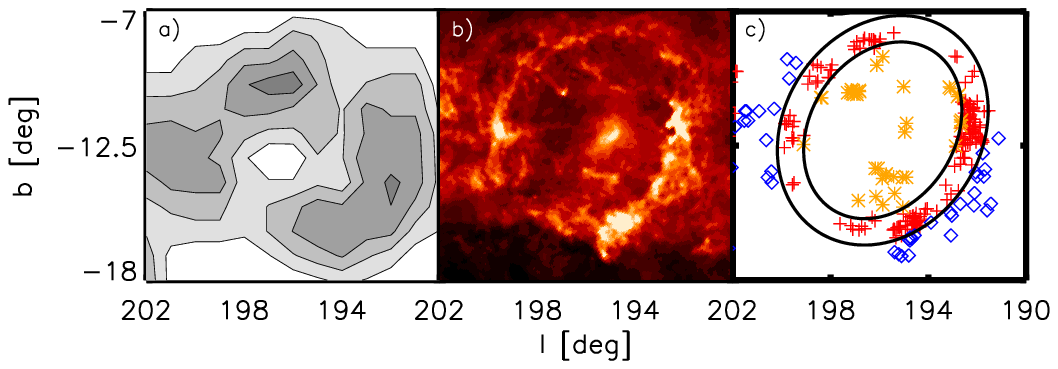}
   \caption{The $\lambda$ Orionis region. Figure a): grayscale contour map showing the N/W(CO) values, contour levels are at 3.2, 10, 32 and 100 $(Kkms^{-1})^{-1}$. Figure b): 100 $\mu$m surface brightness of the same region from the I100 map of Schlegel et al. (1998). Figure c): sources classified as YSO candidates in the same region. Orange asterisks present QDA YSOs located IN the loop, red crosses are for sources ON the loop, and blue diamonds present those candidates located in OFF regions. Black ellipses present the borders of the GIRL G195-11. The surface density of the YSO candidates shows an excess on the loop compared to the environment.}
              \label{YSOratio}
\end{figure*}

We compared $N$ to the $^{12}$CO line intensity W(CO) [Kkms$^{-1}$] of Dame et al. (2001) on a 1 degree scale. We found that all the major star forming regions, like Orion, TAP region, Polaris Flare, Cepheus Flare, Lupus, Vela sheet, Cyg X, Cyg OB7, Carina and Aquila Rift show excess (see top and middle panel in Fig. \ref{activeloopouter}). The spatial distribution of the excess seems to be structured. An example is the expanding $\lambda$-Orionis Molecular Ring (Maddalena \& Morris 1987, Lang et al. 2000) around the O8 star $\lambda$ Ori (see Fig. \ref{YSOratio}/a). It is one of the most active star forming regions, $10[Kkms^{-1}]^{-1}<N$/W(CO)$<100[Kkms^{-1}]^{-1}$ (see Fig. \ref{activeloopouter}). The ring was probably produced by a SNe II (Dolan \& Mathieu 2002), triggered star formation on the ring was reported by Lee et al. (2005).\\

\subsubsection{IR loops and YSOs}

We also studied the correlation in the celestial position of the QDA YSOs with those of the galactic FIR loops (GIRLs, K\"{o}nyves et al. 2007). These structures are characterised by an underdensity or excess density of interstellar matter and are thought to be directly connected to the star-formation process (Blaauw 1991), forming loop-like, hole-like and filamentary-like structures. GIRLs, by definition must show an excess FIR intensity confined to an arc-like feature extending to at least 60\% of a complete ellipse-shaped ring (see Fig. \ref{YSOratio}/b).

Fig. \ref{YSOratio}/c shows an example of the QDA YSO distribution relative to the loop borders. Orange asterisks, red crosses and blue diamonds present the sources located IN, ON and OFF relative to the loop, respectively. The surface density of sources over the averaged E(B-V) value for the given area ($N$/E(B-V)) shows a significant excess ON the GIRLs at moderate galactic latitudes (i.e. at $3^{\circ}<|b|<28^{\circ}$). We chose these regions, because the reliability of the GIRLs is lower in the mid plane because of the high confusion, while at high galactic latitudes we have almost no YSOs. We note also that the nature of loops at high and low galactic latitude is also different (see K\"{o}nyves et al. 2007).

For detailed $N$/E(B-V) values see Table ~\ref{mc}. To test the randomness of the observed excess we obtained 500 MC simulations. In each of the 500 MC samples the observed number of sources were randomly placed onto the sky following the same marginal distributions in galactic longitude (l) coordinates with a resolution of 5 degree. Along the galactic latitude (b) we defined ranges associated with the quantiles of the distribution (10\%, 20\%, ..., 90\%) and sources were placed to follow these ranges. Results of the statistical analysis are shown in Table ~\ref{mc}, where columns are as follows: (1) Position relative to the GIRLs, (2) Observed $N$/E(B-V) values, (3) Average $N$/E(B-V) value of the 500 MC simulations, (4) Standard deviation of the simulated values, (5) minimum value and (6) maximum value in the simulations. To find out if the observed excess ON the loops is significant, we performed a one sample $t$-test. By conventional criteria, the difference between the observed and the simulated values is considered to be statistically significant.

\begin{table}[!ht]

\caption{All-sky and outer galaxy comparison of observed and simulated $N$/E(B-V) values for the IN, ON and OFF regions.}
\label{mc}   
\centering
\begin{tabular}{lccccccc}
\hline                  
&Region& Obs. &$\overline{MC}$ & $\sigma$ & $MC_{-}$ & $MC^{+}$\\
\hline\hline
&IN & 0.95 & 1.22 & 0.04 & 1.11 & 1.34\\
All-sky&ON & 2.06 & 1.19 & 0.03& 1.13 & 1.27\\
&OFF & 1.33 & 1.58 & 0.01 & 1.54 & 1.61\\
\hline        
&IN & 1.08 & 1.27 & 0.04 & 1.13 & 1.39\\\
Outer &ON & 2.44 & 1.28 & 0.03& 1.19 & 1.37\\
Galaxy&OFF & 1.02 & 1.75 & 0.02 & 1.67 & 1.82\\
\hline        
                        
\end{tabular}
\end{table}

Since the significance of the result is very high we tried to establish if the observed excess is due to a general excess or it is caused by a few very active loops. Therefore we carried out a loop-by-loop analysis. First we counted the number of sources seen ON for each individual loops ($n$). This $n$ was then divided by the measured area ($n/A=N$) and the measured average E(B-V) value for the given loop. This way we calculated the $N$/E(B-V)$_k$ value for all the $k$ loops. To have an estimation of the $N$/E(B-V) value fluctuation, in each of the 500 MC simulations and for each of the $k$ loop the $N$/E(B-V) values were calculated. Finally, from the MC simulations we calculated the mean $N$/E(B-V)$_k$ for all the $k$ loop ($\overline{N/\rm{E(B-V)}}_k$) along with their standard deviation $\sigma_k$. 

\begin{equation}
\overline{N/\rm{E(B-V)}}_{k}=\sum_{i=1}^{500}(N/\rm{E(B-V)_{i,k}})/500
\end{equation}

Our QDA YSOs were found to be associated with 98 loops. After the detailed loop-by-loop analysis 43 of the 98 were found to have $N/\rm{E(B-V)}_k > \overline{N/E(B-V)}_k+3\sigma_k$. 

\begin{figure}[!ht]
   \centering
	\includegraphics[width=9cm]{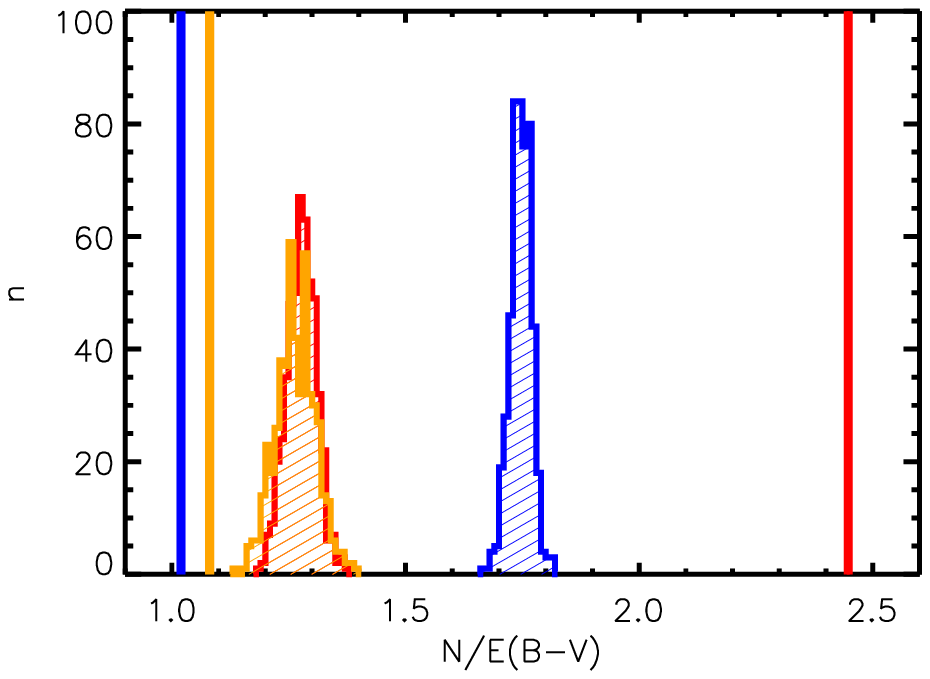}
   \caption{Histograms: distribution of $N$/E(B-V) values in 500 MC simulated samples. Vertical lines show the measured $N$/E(B-V) values for the QDA selected sample. Red, orange and blue colors correspond to the ON, IN and OFF regions, respectively. Investigated area was $90^\circ<l<270^\circ$ and $3^\circ<|b|<28^\circ$. The observed excess for the ON regions is  1.16, while the observed shortage for the IN regions is 0.19.}
              \label{outerelosav}
\end{figure}

\begin{figure}[!ht]
   \centering
	\includegraphics[width=8cm]{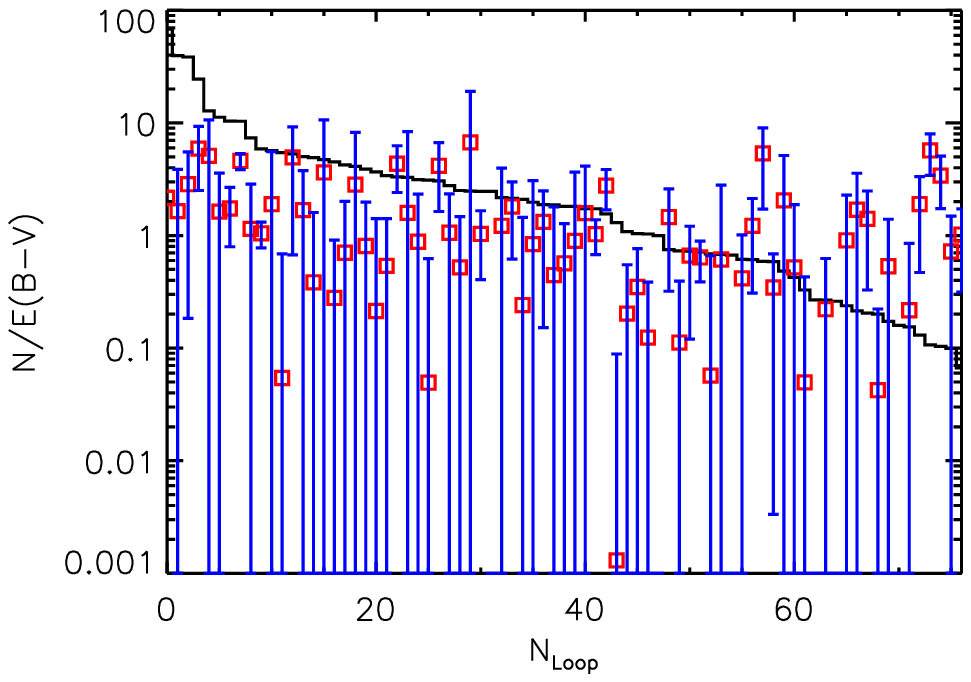}
   \caption{One-by-one activity analysis of loops located in the outer Galaxy ($90^\circ<l<270^\circ$ and $3^\circ<|b|<28^\circ$). Black solid lines present the observed relative surface density $N/\rm{E(B-V)}_k$ values in decreasing order. Red squares present the mean simulated relative surface density $\overline{N/\rm{E(B-V)}}_k$ values for given loops with the corresponding $3\sigma$ error bars plotted as blue lines.}
              \label{szignifikansloopouterhistogram}
\end{figure}

\begin{figure*}[!ht]
\centering
\begin{tabular}{c}
	\includegraphics[trim=-1cm 0cm 0.5cm 6cm, width=0.8\textwidth]{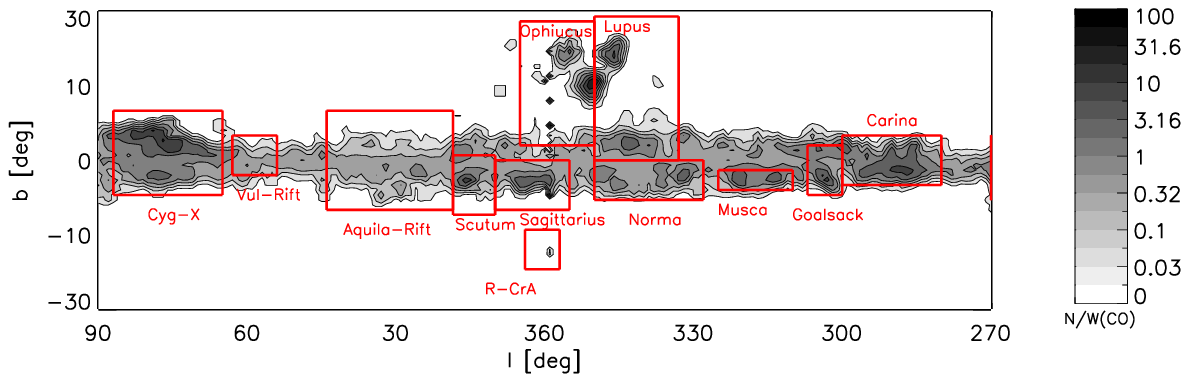}\\
	\includegraphics[trim=-1cm 0cm 0.5cm 6.5cm, width=0.8\textwidth]{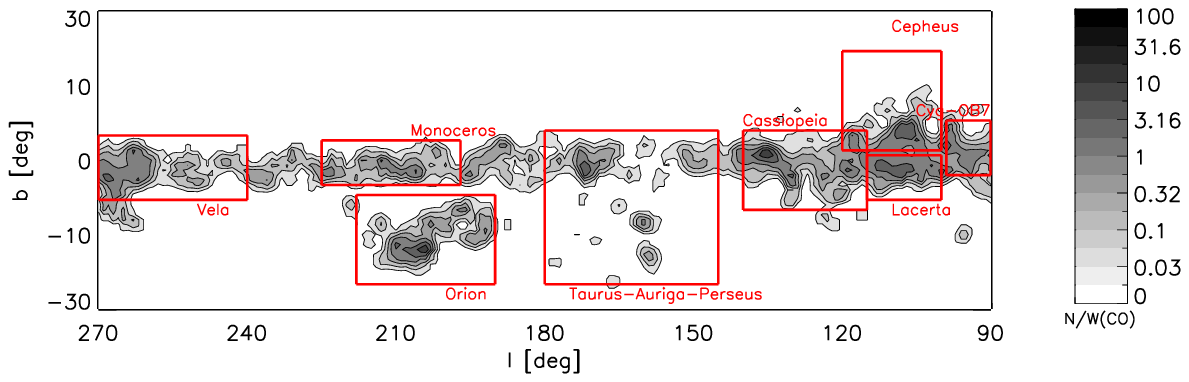}\\
	\includegraphics[trim=-1cm 0cm 0.5cm 6.5cm, width=0.8\textwidth]{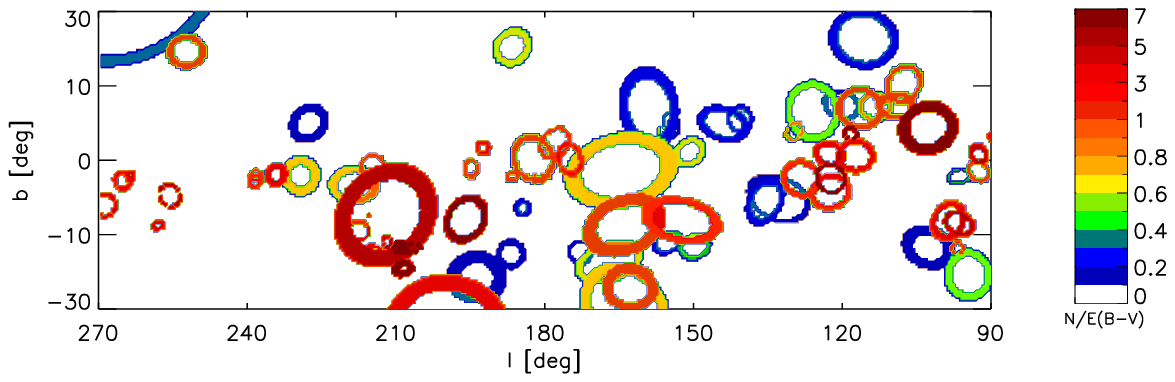} 
\end{tabular}

\caption{{\bf Top panel:} The ratio of N and W(CO) of Dame et al. (2001) on a smoothed 2 degree scale shown for the inner regions of the Galaxy ($90^{\circ}>l>270^{\circ}$). The contour lines correspond to the N/W(CO) values indicated with the colour bar. {\bf Middle panel:} Same as top panel, shown for the outer Galaxy ($90^{\circ}<l<270^{\circ}$). {\bf Bottom panel:} Activity colour map of the 77 active GIRLs located in the outer Galaxy and centered in the $3^\circ<|b|<28^\circ$ region. The colour code corresponds to the GIRLs activity measured as averaged $N$/E(B-V) on each loop.}
\label{activeloopouter}
\end{figure*}

We also noticed that 77 loops of the 98 are located in the outer Galaxy ($90<l<270$) (see Fig. \ref{activeloopouter}). This is partly caused by a selection effect. The detection of the GIRLs was better in the outer regions than in the direction of the central parts of the Galaxy (K\"{o}nyves et al. 2007). In the same way as for the all-sky we carried out an analysis to see the star formation activity on those loops. For the $N$/E(B-V) values calculated for the integrated IN, ON and OFF regions see Table \ref{mc} and Fig. \ref{outerelosav}. Significance tests for the loop activity were carried out in a similar fashion. For the significance histogram see Fig. ~\ref{szignifikansloopouterhistogram}. We found that 40 of the 77 loops are truly active. 

To differentiate between YSO located ON, IN and OFF, the WISE and AKARI colours of YSOs were investigated as a function of position relative to the loops. We found, that the WISE W1-W2, and AKARI [F65/F90], [F90/F140] and [F140/F160] colours are very similar in all positions, and the difference of means is always within the standard deviation. We also tried to differentiate by using QDA. The result was very unreliable, suggesting that YSOs have similar colours independently from their position. \\

\subsubsection{Cold cores in association with YSOs}

On the smallest scale we investigated the correlation between the QDA YSOs and the cold ISM clumps listed in the Planck Early Cold Core (ECC, Planck Collaboration, 2011) sample. ECC contains 915 highly reliable detections as a part of the Planck Early Release Compact Source Catalog (ERCSC), with colour temperatures below 14 K. We searched YSOs in the vicinity of cold cores with radius values equal to the given ECC's semi-major axis, major axis, and 2 times the major axis. The observed correlation was tested again by using the same MC simulations as before. We found that significantly more YSOs are located in the vicinity of ECC objects that one expects based on the simulated samples. Also, we observed significantly more ECCs having nearby YSOs than the predicted values. For details see Table ~\ref{eccysotableallsky}. Columns are as follows: 1) radius used for finding associated YSOs; 2) total number of YSOs found to be associated with ECCs within the given radius; 3) average of the total number of YSOs found within the given radius in the MC simulations; 4) total number of ECCs found to be associated with 1 ore more YSOs; 5) average of the total number of ECCs in the simulated samples; 6) average number of YSOs in the given vicinity of those ECCs with 1 or more YSOs; 7) average calculated from the simulated sample; 8) the maximum number of YSOs associated with a given ECC; 9) average of the maximum number of simulated YSOs associated with a given ECC in the MC simulations.

\begin{table}
\caption{Number of YSOs associated with ECCs in the QDA sample and number of ECCs having at least one associated QDA YSO on the all-sky. Reliability of results was tested with Monte-Carlo simulations (see text).}
\label{eccysotableallsky}
\centering
\begin{tabular}{c|cc|cc}
\hline
r & \multicolumn{2}{c}{$N_{\star}$} & \multicolumn{2}{c}{$N_{ECC}$} \\
$[amin]$ & obs & MC & obs & MC \\
\hline
\hline
		SMA		&	213	&	6.41$\pm$2.49		&	163	&	6.26$\pm$2.41	\\
		MA		&	470	&	25.58$\pm$5.26	&	259	&	23.66$\pm$4.72\\
		2xMA	&	1255	&	102.62$\pm$11.26	&	379	&	79.19$\pm$8.41\\

\end{tabular}

\end{table}

\subsection{Comparison to existing catalogues}

Pollo et al. (2010) separated the galaxies and stars in the $\beta-1$ version of the AKARI FIS  BSC by using 5176 objects selected from the low-extinction regions of the Galaxy. Their selection criteria did not include any flux quality restrictions. For the separation they used colour-flux and colour-colour diagrams, all combinations constructed from the 4 different FIS wavelength band flux densities. They found two "clouds" in the colour-flux and colour-colour spaces. One cloud contained galaxies in 95\% while the other had stars in more than 80\%. They noted that among the stars which are overlapping with the galaxy-cloud they found a few T Tauri stars. Our analysis included 13 times more point sources with the best flux quality in the W90 and W140 bands. We found that on average 41.2\% of the objects we identified as YSO candidates matches their criteria for galaxies. At the same time, in average 32\% of the objects we identified as galaxies belong to their galaxy cloud on the colour-colour diagrams.

Gutermuth et al. (2009) classified 2548 YSOs using Spitzer IRAC and MIPS. The whole sample of FIS sources we used for the selection has an overlap with theirs in 43 cases, while our QDA YSO selection has an overlap with theirs in 39 cases by using a 30$^{\prime \prime}$ search radius. As a result, 4 (9.3\%) of their classified sources was misclassified with QDA.

Rebull et al. (2011) investigated the Taurus-Auriga region for YSO candidates from a $\approx$260 sqd region, based on 2MASS+WISE data. They identified 196 previously known YSOs and 94 new YSO candidates, altogether with 686 confirmed galaxies and some other objects. The AKARI FIS BSC subsample we used in this study has an overlap with their objects in 83 cases, using 30$^{\prime \prime}$ search radius. 35 out of the 83 is QDA classified YSO candidate. Among the 35 sources they found only 1 was a galaxy, based on the SED shape of the source. Another source was found to be a planetary nebula in their study. The remaining 33 QDA YSOs was found to be known YSO including one newly classified. 48 of the 83 cases were not classified as QDA YSO candidates. Based on their results, only 2 of the 48 are known YSOs, while the remaining 46 are galaxies. As a result, the misclassification rate among the QDA YSOs is  5.7\%, and we missed 4.1\% of the known YSOs in this region.

Another study analysed the YSO content of the Taurus-Auriga regions. Takita et al. (2010) used the MIR part of the AKARI survey. They reported 133 detections among the 517 known T Tau stars. Our AKARI subsample overlaps with their reported 133 sources in 29 cases. Among the 29 sources we have 26 QDA YSOs. Only 10\% of the possible matches has been misclassified.

Koenig et al. (2008) lists details for 17771 sources from the Spitzer IRAC and MIPS observations in and around the W5 HII region. From the 96 sources existing in our AKARI subsample, 91 were found to be QDA YSOs. This means that only 5.2\% of the possible sources have been misclassified.

Harvey et al. (2007) lists 286 YSOs and YSO candidates for the Serpens cloud. The studied AKARI subsample has an overlap with their sources in 7 cases. 6 of them are also QDA YSOs, and only 1 has been missed. 

Povich et al. (2011) presented a catalogue of 1439 YSOs surveyed in a 1.42 sqd region by the Chandra Carina Complex Project. The studied AKARI subsample overlaps with their catalogue in 33 cases. All the 33 sources are also QDA YSOs.

Kang et al. (2009) selected 737 candidate embedded YSOs from Spitzer Space Telescope data in the W51 giant molecular cloud region. From the overlapping 27 AKARI subsample sources we found 26 QDA YSOs; only one source was missed. 

Kun et al. (2009) found 77 pre-main sequence stars in the Cepheus flare region. 100\% of the 12 overlapping sources were identified as QDA YSOs.

Felli et al. (2002) selected a total of 715 YSO candidates from the ISOCAM observations during the ISOGAL program. The overlap between their sample and the AKARI subsample was 44 sources. 43 of these objects are also QDA YSOs. 

Minier et al. (2003) searched for 6.7 GHz methanol masers toward low-mass YSOs. All of the 12 overlaps were identified as QDA YSOs.

A quick classification of YSOs may be possible using fewer photometric bands (see e.g., WISE based selections by Koenig et al. 2011, and by Marton et al. in prep.). However, our YSO catalogue is a better input for detailed analysis of YSOs since it includes YSOs with FIR photometric data. We compared the galaxy contaminations in a comparison with the Koenig et al. (2011) method. Following their description we looked for extragalactic objects, shock emission blobs and PAH emission objects among the QDA YSO sample. We flagged 3935 objects as PAH/star forming galaxy, 5880 objects as AGN candidates, 599 sources as shock objects and 7954 as PAH emission objects. We looked for these flagged sources in the SIMBAD database following the same method as described before. We found that 6.8\% of the flagged sources belong to those we used to identify YSOs in the teaching phase. This is slightly lower than that 8.6\% found in our QDA YSO sample.

Recently, Majaess (2012) described a spectral index ($\alpha$) classification scheme, based on 2MASS and WISE data. He reported the detection of $n>10^4$ YSO candidates. We applied this $\alpha$ classification scheme on our QDA YSOs. We found 22594, 7905, 8872 and 4630 sources fit the criteria of Class I, Flat, Class II and Class III objects, respectively. The reported number of Class I/F objects is $30\times10^3$, which is in very good agreement with our results (30499).

ATLASGAL is the APEX Telescope Large Area Survey of the Galaxy (Schuller et al. 2009), an observation programme that uses the LABOCA bolometer array at APEX. This survey aims at mapping over 400 square degrees at 870 microns in the inner Galaxy, with a uniform sensitivity of a few solar masses at 1 kpc distance (see e.g. Tackenberg et al. 2012, Contreras et al. 2013, ). 18806 candidate AKARI YSOs are located on this area. The GLIMPSE survey (Benjamin et al. 2003) of the Spitzer Space Telescope covers roughly the same area as the ATLASGAL survey. We found 4219 candidate YSOs which have no associated Spitzer point source within 15$^{\prime \prime}$. We found that 22 $\%$ of the AKARI candidate YSOs were not observable with the SST.

Correlation of YSO candidates and IR loops on the all-sky with such a high number of sources was carried out for the first time. Kiss (2007) selected 3872 classical T Tauri star candidates from the 2MASS PSC and studied their correlation with the GIRLs in the $2^{nd}$ and $3^{rd}$ galactic quadrants. Kiss found an excess in the CTT density distribution on the GIRL shells and statistically proved that this excess does not exist by chance. This result is in very good agreement with our results.

Observations and theory suggest that not every bubble is capable of triggering star formation, because the effect depends on various independent environmental and initial conditions. Detailed discussion on triggered star formation in expanding bubbles or shells can be found in Elmegreen (1998) and Elmegreen (2011). We found that not all GIRLs are sites of active star formation. 271 GIRLs are located in the analysed region of the sky, 173 (65\%) of them are located in the outer Galaxy ($2^{nd}$ and $3^{rd}$ quadrants). 98 (36\%) was found to be star forming loops, 77 (28\%) are located partly or entirely in the outer Galaxy. The number of loops that are actively forming stars was found to be 43 (16\%). 40 of them (15\% of all) are located in the outer Galaxy. Thompson et al. (2012) analysed the overdensity of massive YSOs around 322 mid-infrared Spitzer bubbles. They found that about a quarter of the bubbles show signs of triggered star formation based on the YSO density on and around the bubbles. They also found that this percentage is in good agreement with former and similar studies of Deharveng et al. (2010) and Watson et al. (2010). These studies analysed the correlation of YSOs and bubbles located inside the galactic midplane. Therefore our study can be considered as a complementary analysis, since we investigated the correlation of YSO candidates and loop-like structures at moderate galactic latitudes. The 15\% efficiency is somewhat lower than the aforementioned values, but one could expect a lower value, given the more relaxed conditions and the lower ISM pressure above the midplane. The surfaces of high and low density ISM will be better revealed by the Planck all-sky survey.

Studying the correlation between the Planck ECC clumps (Planck Collaboration, 2011) and the QDA YSOs revealed that 163 of the 915 clumps (17.8\%) have at least one associated QDA YSO within a radius equal to the semi-major axis of the clump. 259 (28.3\%) have QDA YSOs within a region equal to the major axis and 379 (41.4\%) have QDA YSOs within the double of the major axis. We found that the temperature of the clumps is not significantly different whether they have associated YSOs or not, but all clumps having at least one associated YSO are warmer than 6.7 K. Wu et al. (2012) observed CO toward 674 ECC clumps. We used their results in order to find out if there is a correlation between the derived physical properties and the star formation activity of those clumps. The N(H$_2$) values are similar in the clumps independent of the number of the associated YSOs. We identified a limit value of 1.6$\times 10^{22}$ cm$^{-2}$, above which all clumps seem to form new stars within the radius of double the major-axis.

\section{Summary}

We identified 44001 AKARI FIS BSC sources as YSO candidates. The vast majority of these sources are from Class I type according to the Lada classification. For these sources (i.e. with MIR and FIR flux density data) the physical parameters of the cold circumstellar material can be also derived. Thus our catalogue may be the basis of further detailed studies of YSOs. Their large-scale distribution was investigated comparing the YSO surface density to the distribution of the galactic CO of Dame et al. (2001). All the major SFRs showed a high concentration of YSOs compared to the CO line intensity, appearing as an excess in the  $N$/W(CO) ratio. On medium-scale, the distribution of YSOs was compared to that of the galactic IR loops (GIRLs, Kiss et al. 2004,  K\"{o}nyves et al. 2007). We found that the $N$/E(B-V) values show a clear excess on the GIRLs in the outer galaxy ($90^\circ<l<270^\circ$ and $3^\circ<|b|<28^\circ$). We tested the correlation by comparing the results to those coming from Monte-Carlo simulations. We found the result to be highly significant, based on the performed one sample $t$-test. 
Our loop-by-loop analysis showed that YSOs are associated with 98 loops. 43 of them were found to have a significant excess star formation. 40 of them are located partly or entirely in the outer Galaxy. MIR and FIR colours of YSOs are independent from their position relative to the FIR loops. 
Significant excess of AKARI YSOs were found ECCs based on Monte-Carlo simulations. A critical column density for star formation is 1.6$\times 10^{22}$ cm$^{-2}$. 

%The catalogue and description of AKARI FIS galaxies (see Section \ref{results}.) will be presented in our forthcoming paper. \\

\textit{Acknowledgement}: We would like to thank T. J. Craven, L. D. Krista and our anonymous referee for their valuable comments and corrections. The work of G.M. has been supported by the following grants: i) PECS contract no. 98073 of the Hungarian Space Office and the European Space Agency, ii) Hungarian Research Fund (OTKA) grant no. K104607. The project was also supported by the OTKA K101393. The European Union and the European Social Fund have provided financial support to the project under the grant agreement no. T\'AMOP-4.2.1/B-
09/1/KMR-2010-0003. This research is based on observations with AKARI, a JAXA project with the participation of ESA. This paper was resulted in a JSPS-HAS coo-funded collaboration project. This research has made use of the VizieR catalogue access tool, CDS, Strasbourg, France. This research has made use of the SIMBAD database, operated at CDS, Strasbourg, France. This publication makes use of data products from the Wide-field Infrared Survey Explorer, which is a joint project of the University of California, Los Angeles, and the Jet Propulsion Laboratory/California Institute of Technology, funded by the National Aeronautics and Space Administration.

%Acknowledgement should be placed at end of main text.
%(NOT after the Appendix.)

%\appendix
%\section*{Complete data}

%%%
% See the manual for the detail.
%%%

\end{document}